# SPH Simulations of Accretion Disks and Narrow Rings

Sarah T. Maddison[1]*, James R. Murray[2]† and Joe J. Monaghan[1]
[1]Maths Department, Monash University, Clayton, 3168, Australia
[2]CITA, University of Toronto, Ontario, M5S 1A7, Canada

June 6, 1995

## Abstract

We model a massless viscous disk using Smoothed Particle Hydrodynamics (SPH) and note that it evolves according to the Lynden-Bell & Pringle theory (1974) until a non-axisymmetric instability develops at the inner edge of the disk. This instability may have the same origin as the instability of initially axisymmetric viscous disks discussed by Lyubarskij et al. (1994). To clarify the evolution we evolved single and double rings of particles. It is actually inconsistent with the SPH scheme to set up a single ring as an initial condition because SPH assumes a smoothed initial state. As would be expected from an SPH simulation, the ring rapidly breaks up into a band. We analyse the stability of the ring and show that the predictions are confirmed by the simulation.

# 1 Introduction

As a test case for a two dimensional SPH code (for an overview of SPH, see for example Monaghan 1992) developed to study the viscous evolution of accretion disks we modelled the spreading of a narrow viscously shearing ring of matter. We wished to compare simulations with the results of thin disk theory as developed by Lynden-Bell and Pringle (1974). Lubow (1991) used the rate of spread of an axisymmetric ring of viscously interacting SPH particles to estimate the shear viscosity due to the artificial viscosity term in his code. Flebbe et al. (1994) tested a tensor form of a general Navier Stokes viscosity term by comparing an axisymmetric ring simulation with Lynden-Bell and Pringle's result for the evolution of a $\delta$ function density distribution. Because SPH cannot be used with a $\delta$ function density, Flebbe et al. used an approximate initial condition. We prefer to be consistent and simulate a configuration with an initial state which can be resolved by SPH. In particular we looked at an axially symmetric annulus with an initial surface density

$$\Sigma_0(r) = \begin{cases} \exp\left\{-\left(\frac{r-r_o}{l}\right)^2\right\} & r_1 < r < r_2 \\ 0 & \text{elsewhere} \end{cases} \quad (1)$$

where $r_o$ and $l$ are the radius of maximum density and width of the Gaussian respectively. The annulus is assumed to be Keplerian with the only forces considered being the gravitational attraction of the central object and bulk and shear viscosity forces within the annulus. Pressure and disk self-gravity are not considered. To simplify the analysis a constant kinematic viscosity $\nu$ was assumed. For the above initial condition the general solution for the surface density at radius $r$ and time $t$ is given by

$$\Sigma(r,t) = \frac{1}{r^{1/4} 6\nu t} \int_{r_1}^{r_2} r'^{5/4} \exp\left\{-\left(\frac{r'-r_o}{l}\right)^2\right\}$$
$$\times \exp\left\{-\frac{(r^2+r'^2)}{12\nu t}\right\} I_{1/4}\left(\frac{r'r}{6\nu t}\right) dr'$$

where $I_{1/4}(x)$ is a modified Bessel function of the first kind. For large arguments, $I_{1/4}(x) \approx \exp(x)/\sqrt{2\pi x}$. Thus for small $t$ we have the approximate solution:

$$\Sigma(r,t) \approx \frac{1}{r^{3/4}\sqrt{\pi 12\nu t}} \int_{r_1}^{r_2} r'^{3/4} \exp\left\{-\left(\frac{r'-r_o}{l}\right)^2\right\}$$
$$\times \exp\left\{-\frac{(r'-r)^2}{12\nu t}\right\} dr' . \quad (2)$$

*maddison@hypatia.maths.monash.edu.au
†jmurray@cita.utoronto.ca

Following the equations of Pringle (1981), the simulation is given an initial radial velocity

$$v_r = -\frac{3}{\Sigma R^{1/2}} \frac{\partial}{\partial R}\left[\nu \Sigma R^{1/2}\right]$$

where $\Sigma$ is our initial Gaussian.

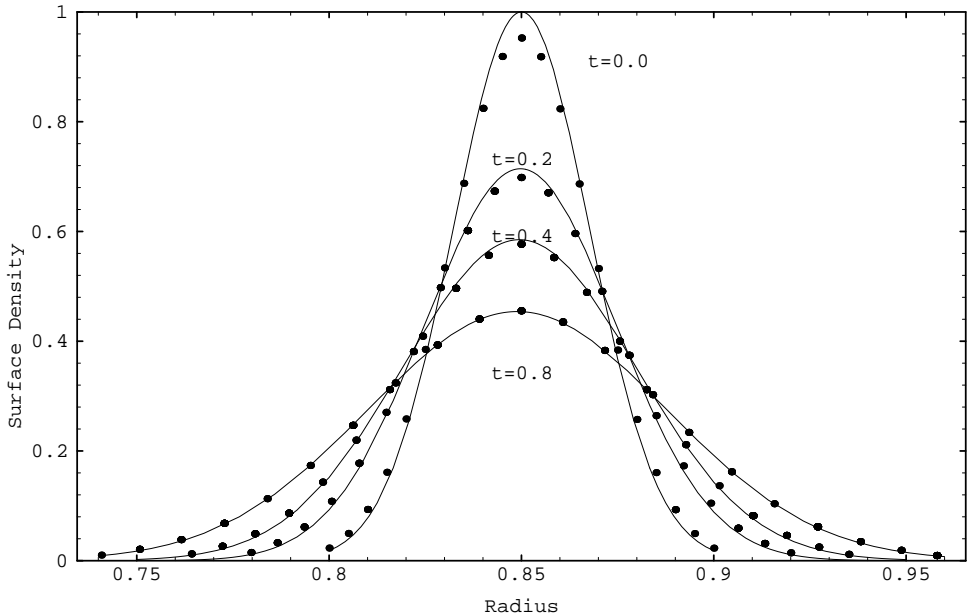

Figure 1: The time evolution of the surface density of a Gaussian disk. The solid lines denote the theoretical solution at the times shown. The heavy points show the corresponding SPH results. For this simulation we used $r_1 = 0.80$, $r_0 = 0.85$, $r_2 = 0.90$, $l = 0.025$ and $\nu = 2.5 \times 10^{-4}$.

The theoretical and simulated spread of an annulus with a Gaussian density profile is shown in figure 1. The best results were obtained when the particles were set up in concentric rings so as to give a uniform number density throughout the annulus. The Gaussian profile was set up by giving each particle a mass proportional to $\Sigma_0(r)$. In this particular case 22,420 particles in 21 rings were used (see figure caption for further parameters of the model).

After ten revolutions of the particles at $r = r_1$ the inner edge of the disk started to break up. Other simulations of disks made up of more rings gave similar results. After several rotation periods, the viscosity spread the rings to the point where the innermost ring became somewhat separated from the remaining rings. It then started to lose shape and break up. This effect marched its way through the remaining disk. One would like to know if this is a genuine instability or an artifact of the numerical method. In the following we give a partial answer by analysing the stability of a single ring.

# 2  Stability Analysis

We consider the stability of a single viscous ring of equi-spaced SPH particles in orbit with angular velocity $\Omega$ around a central mass $M$. Again, gravitational and pressure forces of the ring particles have been neglected to investigate the viscous forces.

Each particle $i$ has unperturbed and perturbed positions in radial and azimuthal coordinates:

$$\begin{aligned}
(r_o, \phi_o) &= (a, \Omega t + \Theta_i) \\
(r, \phi) &= (a(1+q_i), \Omega t + \Theta_i + \sigma_i)
\end{aligned}$$

where $a$ is the ring radius, $\Omega t$ is a rotating reference line, $\Theta_i$ is the unperturbed azimuthal position, and $\sigma_i$ and $aq_i$ are the azimuthal and radial perturbations respectively.

The radial and azimuthal equations of motion governing a particle in the ring are:

$$\frac{d^2 r}{dt^2} - r\left(\frac{d\phi}{dt}\right)^2 = -\frac{\mathcal{G}M}{r^2} + F_r \tag{3}$$

$$r\frac{d^2\phi}{dt^2} + 2\frac{dr}{dt}\frac{d\phi}{dt} = F_\phi , \tag{4}$$

where $\mathbf{F} = (F_r, F_\phi)$ is the viscous force per unit mass.

Using the length and time scales $a$ and $t = \tau\sqrt{a^3/(\mathcal{G}M)}$ we find:

$$\frac{d\phi}{d\tau} = 1 + \frac{d\sigma}{d\tau} ,$$

and, by linearising equations (3) and (4),

$$\frac{d^2 q}{d\tau^2} - 3q - 2\frac{d\sigma}{d\tau} = \frac{a^2}{\mathcal{G}M}F_r \tag{5}$$

$$\frac{d^2\sigma}{d\tau^2} + 2\frac{dq}{d\tau} = \frac{a^2}{\mathcal{G}M}F_\phi . \tag{6}$$

For the *non*-viscous case, where $F_r = F_\phi = 0$,

$$\ddot{q} - 3q - 2\dot{\sigma} = 0 \tag{7}$$

$$\ddot{\sigma} + 2\dot{q} = 0 . \tag{8}$$

We consider perturbations of the form:

$$q = A\exp\{i(k\Theta + \omega\tau)\} \tag{9}$$

$$\sigma = B\exp\{i(k\Theta + \omega\tau)\} , \tag{10}$$

where $\omega$ is the frequency of the disturbance and $k$ is the wave number. Substituting (9) and (10) into the linearised equations (7) and (8), we find that $\omega = 0, \pm 1$, with a double root at $\omega = 0$. The general solution for this non-viscous case is the superposition of (a) an oscillation, (b) a shift in $\Theta_i$ without changing $a$, and (c) the motion associated with all particles being shifted to a different circular orbit (see Murray 1994).

Returning now to the *viscous* case, we need the SPH viscous forces. Assuming that the SPH kernel, $W_{ij}$, is Gaussian (normalised for two dimensions) we can use:
$$\nabla_i W_{ij} = -\frac{2\mathbf{r}_{ij}}{h^2} W_{ij},$$
where $h$ is the SPH smoothing length and $\mathbf{r}_{ij} = \mathbf{r}_i - \mathbf{r}_j$. Substituting this into the formula for the viscous force per unit mass (Monaghan 1992, eqn 4.1) and using $\rho_i = \epsilon m_i/h^2$, where $\rho_i$ and $m_i$ are the density and mass respectively of particle $i$, and $\epsilon$ is a positive dimensionless constant of order unity that depends on the spacing of the particles,
$$\begin{aligned}\mathbf{F}_i &= \sum_j \frac{m_j}{\rho_j} \alpha \bar{c}_{ij} h \left[\frac{\mathbf{v}_{ij}.\mathbf{r}_{ij}}{r_{ij}^2 + \eta^2}\right] \nabla_i W_{ij} \\ &= -\frac{2\alpha c h}{\epsilon} \sum_j \left[\frac{\mathbf{v}_{ij}.\mathbf{r}_{ij}}{r_{ij}^2 + \eta^2}\right] \mathbf{r}_{ij} W_{ij},\end{aligned}$$
where $\alpha$ is the viscosity coefficient, $c$ is the speed of sound (replacing $\bar{c}_{ij}$ since the sound speed is constant in the isothermal case) and $\eta$ prevents singularities in the viscous force for very small particle separation. We can ignore $\eta$ for the purpose of this stability analysis. The kinematic viscosity is given by $\nu = \alpha h c/8$.

Since we are considering the linearised equations, we need only consider perturbations to $\mathbf{v}_{ij}.\mathbf{r}_{ij}$ since it vanishes in the unperturbed case. After linearisation (see appendix) we find:
$$\begin{aligned}\mathbf{v}_{ij}.\mathbf{r}_{ij} &\sim (\dot{\sigma}_j - \dot{\sigma}_i) \sin \zeta \\ r_{ij}^2 &\sim 2(1 - \cos \zeta),\end{aligned}$$
where $\zeta = \Theta_j - \Theta_i$. The scaled radial component of the force per unit mass on particle $i$ is then given by:
$$\frac{a^2}{\mathcal{G}M} F_r = -\frac{g}{2} \sum_j (\dot{\sigma}_j - \dot{\sigma}_i) \sin \zeta \; W_{ij} \;, \qquad (11)$$

and the tangential force per unit mass component on particle $i$ is:
$$\frac{a^2}{\mathcal{G}M} F_\phi = \frac{g}{2} \sum_j (\dot{\sigma}_j - \dot{\sigma}_i)(1 + \cos \zeta) W_{ij} \;, \qquad (12)$$

where $g = 2\alpha ch/\epsilon$.

We now have the full SPH formulation of the equation of motion of a viscous ring, by combining (5) & (6) and (11) & (12), and substituting (9) and (10). The summation is over all $N$ particles in the ring. However, only nearby particles contribute to the sum, which can thus be replaced by an infinite sum. It is then clear that the odd components of these sums must vanish, leaving the equation of motion as:

$$(-\omega^2 - 3)A - 2i\omega B = gB\omega\Psi \qquad (13)$$

$$-\omega^2 B + 2i\omega A = -gBi\omega\Phi , \qquad (14)$$

where

$$\Psi = \frac{1}{2}\sum_j \sin(k\zeta)\sin\zeta\ W_{ij}$$

$$\Phi = \frac{1}{2}\sum_j (1 - \cos(k\zeta))(1 + \cos\zeta)W_{ij} ,$$

and if $k = 0$ both $\Psi = 0$ and $\Phi = 0$, thus $F_r = 0$ and $F_\phi = 0$.

Combining (13) & (14) we get a quartic in the frequency $\omega$, one root of which is $\omega = 0$. This leaves us with:

$$\omega^3 - ig\Phi\omega^2 + (2ig\Psi - 1)\omega - 3ig\Phi = 0 . \qquad (15)$$

In the non-viscous case (i.e. where $\Psi = \Phi = 0$) $\omega = 0, \pm 1$, in agreement with the solution of (7) and (8).

If the viscosity (which is proportional to $g$) is small enough, we expect frequencies close to the unperturbed values. We therefore set $\omega = \gamma, 1 + \gamma$ and $-1 + \gamma$ where $|\gamma| \ll 1$ and the $\gamma$ in each case will be different.

<u>$\omega = 1 + \gamma$</u>

Substituting $\omega = 1 + \gamma$ into equation (15) and keeping only the linear terms (since $|\gamma| \ll 1$) we find:

$$\gamma = ig(2\Phi - \Psi) , \qquad (16)$$

and

$$\exp\{i\omega t\} = \exp\{i(1 + \gamma)t\}$$
$$= \exp\{it - gt(2\Phi - \Psi)\} .$$

For stability we require $2\Phi - \Psi \geq 0$. Substituting for $\Phi$ and $\Psi$ we find that for all values of $k$, $2\Phi - \Psi \geq 0$ (see Murray 1994 for this and the following two cases). Thus the $\omega = 1 + \gamma$ mode is stable.

$\underline{\omega = \gamma - 1}$

In this case we get
$$\gamma = ig(2\Phi + \Psi) \qquad (17)$$
and so $\exp\{i\omega t\}$ shows that for stability we require $2\Phi + \Psi \geq 0$, which holds for all values of $k$. Hence $\omega = \gamma - 1$ is also stable.

$\underline{\omega = \gamma}$

Here we get:
$$\gamma = -3ig\Phi \ . \qquad (18)$$
This gives $\exp\{i\omega t\} = \exp\{3g\Phi t\}$ whose exponent is always positive, and hence the $\omega = \gamma$ is an unstable mode for all values of $k$. Note that in the inviscid solution, the $\omega = 0$ case is associated with an azimuthal shift with radius held constant.

For large wavelengths (large compared to the particle spacing, which is of order $h$, and thus large wavelength corresponds to $hk \ll 1$), we substitute a Gaussian for the kernel and approximate the summation by an integral and find:
$$\Phi \approx \frac{k^2 h}{\sqrt{\pi}\Delta\Theta} , \qquad (19)$$
where $\Delta\Theta$ is the azimuthal particle spacing. Thus:
$$\begin{aligned} \omega &= \gamma = -3ig\Phi \\ &\approx -\frac{6\alpha ic}{\epsilon\sqrt{\pi}\Delta\Theta}(hk)^2 \ . \end{aligned} \qquad (20)$$
Since we are assuming that $|\gamma| \ll 1$, we must have:
$$\frac{\alpha c}{\epsilon\sqrt{\pi}\Delta\Theta} \ll 1 \ ,$$
to be consistent with the analysis. In the models that we ran, particles were equally spaced a scaled distance $\Delta\Theta \approx 2\pi/200$ apart, with $\alpha c \ll 1$ (see below for details). Hence, we have found that there is an unstable mode when $\omega \ll 1$, ie when $\alpha c \ll 1$.

The time scale of the viscous instability for long wavelengths is:
$$\tau_\nu \sim \frac{\epsilon\sqrt{\pi}\Delta\Theta}{6\alpha c} \ .$$
In the models we ran, $\alpha = 0.07$, $c = 0.05$ and $\epsilon \approx 4/9$. Thus $\tau_\nu \approx 1.178$ rotation periods. However, from (20) we see that the most rapid growth occurs for short wavelengths for which the approximation (19) to the summation is invalid. Hence we expect instabilities with wavelength approximately a particle spacing to grow fastest.

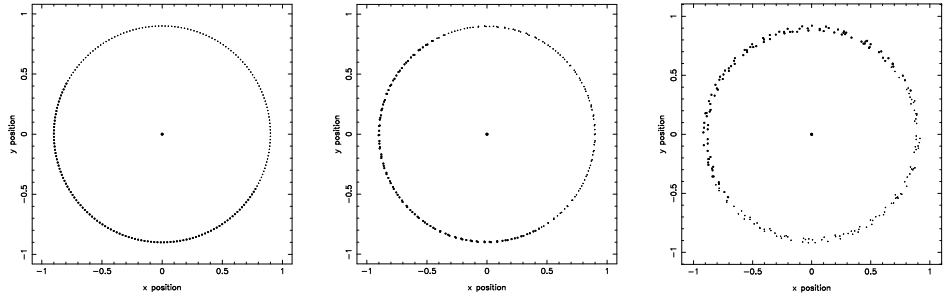

Figure 2: The onset of viscous instability in a single ring system. The figures are at $0.2, 4.1$ and $5.8$ rotational periods.

# 3  The Simulations

We ran an SPH simulation of a single viscous ring of approximately 200 equally spaced particles orbiting a central massive body. The ring was placed at dimensionless radius of 0.9, at which position the ring has a rotational period of 5.4 dimensionless time units. As mentioned above, $\alpha = 0.07$ and $c = 0.05$. Particles started to oscillate very slightly almost immediately and after approximately four rotational periods the ring had broadened. The particles moved on eccentric orbits and after five rotational periods the ring had broadened significantly (see figure 2). It was seen from the simulation that the viscous instability sets in after approximately four rotation periods.

A double ring system was also modelled. If the separation of the rings was less than $2h$ (the radius of influence of the SPH kernel), the two rings acted as one and were disturbed in a similar manner as was the one ring system. However, if the separation of the rings was greater than $2h$, then as usual viscosity acts to spread the rings and particles moved onto eccentric orbits. The inner rings broke up first after four rotation periods, followed by the outer ring at its equivalent rotational phase. The two rings were indistinguishable after eight orbits.

# 4  Conclusions

The results of the disk simulations show that SPH gives results in good agreement with the standard viscous theory. Deviations from theory eventually occur at the inner edge where we observe particles moving from circular orbits. The analysis of Lyubarskij et al. (1994) shows that axisymmetric viscous disks can become unstable whenever the viscous drag on fluid in an elliptical orbit is a maximum at apastron. This is the case for the standard $\alpha$-disk and it is the case for particles in our simulation. For example, the particles in the innermost ring appear to move in and then out which shows that they are individually moving on elliptical orbits with apastron near their neighbouring ring. Since the viscous force is felt when they move near the particles in the neighbouring ring they will be forced into more eccentric orbits.

When modelling viscous disks we found that the innermost ring separated from the remaining disk and then became unstable. We then investigated the dynamics of a single ring. However, the formulation of SPH is based on smoothing. A single ring is the antithesis of a smoothed configuration. The single ring smooths itself by rapidly spreading under the viscous forces. The combination of analysis and simulation confirms the consistency of SPH.

However, the resulting effects of the SPH viscous force prescription in accretion disk simulations is unphysical. The main problem is that we are trying to model a continuum (i.e. a gaseous viscous disk) by a set of discrete rings. For disks made up of concentric rings, this problem can be overcome if the number of rings used is increased so that the rings will not separate due to viscosity by too great a distance (generally less than $2h$). As usual with particle methods, a compromise must be reached between the number of particles and the computation time. Alternatively, one could incorporate a variable smoothing length, so that when the distance between rings is greater than $2h$, the smoothing length increases so that neighbouring rings can still communicate. There are still some problems to be overcome with the implementation of spatially varying smoothing lengths when sharp density gradients are present (see Nelson & Papaloizou 1994).

# A  Appendix

We want to solve for

$$\begin{aligned} s &= \mathbf{v}_{ij} \cdot \mathbf{r}_{ij} \\ &= (\mathbf{v}_i - \mathbf{v}_j) \cdot (\mathbf{r}_i - \mathbf{r}_j). \end{aligned} \quad (A1)$$

Noting that to first order

$$\mathbf{r}_i = (1 + q_i)\hat{\mathbf{r}}_{io} + \sigma_i(\hat{\mathbf{z}} \times \hat{\mathbf{r}}_{io})$$

and differentiating with respect to $t$, (noting $d\hat{\mathbf{r}}_{io}/dt = \hat{\mathbf{z}} \times \hat{\mathbf{r}}_{io}$, with $\Omega = 1$), we find

$$\mathbf{v}_i = (\dot{q}_i - \sigma_i)\hat{\mathbf{r}}_{io} + (1 + \dot{\sigma}_i + q_i)(\hat{\mathbf{z}} \times \hat{\mathbf{r}}_{io}), \quad (A2)$$

where $\hat{\phantom{x}}$ denotes a unit vector, $\mathbf{r}_{io}$ is the unperturbed position vector of particle $i$, and $\hat{\mathbf{z}}$ is perpendicular to the plane of the orbit. We can write the terms contributing to (A1) as, for example,

$$\begin{aligned} \mathbf{v}_i \cdot \mathbf{r}_i &= (\dot{q}_i - \sigma_i)(1 + q_i) + \sigma_i(\dot{\sigma}_i + 1 + q_i) \\ &\simeq \dot{q}_i \\ \mathbf{v}_j \cdot \mathbf{r}_i &= [(\dot{q}_j - \sigma_j)(1 + q_i) + (1 + q_j + \dot{\sigma}_j)\sigma_i]\cos\zeta \\ &\simeq (\dot{q}_j + \sigma_i - \sigma_j)\cos\zeta - (1 + q_i + q_j + \dot{\sigma}_j)\sin\zeta. \end{aligned} \quad (A3)$$

where $\zeta = \Theta_j - \Theta_i$ and $\hat{\mathbf{r}}_{io} \cdot \hat{\mathbf{r}}_{jo} = \cos\zeta$. Similar expressions can be written down for $\mathbf{v}_j \cdot \mathbf{r}_j$ and $\mathbf{v}_i \cdot \mathbf{r}_j$. Combining these expressions we find

$$s = (\dot{q}_i + \dot{q}_j)(1 - \cos\zeta) + (\dot{\sigma}_j - \dot{\sigma}_i)\sin\zeta. \quad (A4)$$

Since $0 < \zeta \ll 1$ we can neglect the first term in (A4) to obtain

$$s \simeq (\dot{\sigma}_j - \dot{\sigma}_i)\sin\zeta.$$